\documentclass[modern]{aastex702}

\usepackage{xcolor}
\definecolor{revisionblue}{RGB}{0,65,180}
\newcommand{\rev}[1]{#1}
\graphicspath{{figures/}}

\newcommand{\msun}{M_\odot}
\newcommand{\kms}{\,\mathrm{km\,s^{-1}}}
\newcommand{\fscat}{f_{\mathrm{scat}}}
\newcommand{\chidof}{\chi^2/\nu}

\shorttitle{3D Electron Scattering in NGC 3783}
\shortauthors{Wada \& Vander Meulen}

\begin{document}
\hypersetup{hidelinks}

\title{Broad Line Regions behind a Haze II: A Three-dimensional Electron-scattering Model of the Broad Br$\gamma$ Line in NGC 3783}

\author[0000-0002-8779-8486]{Keiichi Wada}
\affiliation{Graduate School of Science and Engineering, Kagoshima University, 1-21-35 Korimoto, Kagoshima 890-0065, Japan}
\affiliation{Research Center for Space and Cosmic Evolution, Ehime University, 2-5 Bunkyo-cho, Matsuyama 790-8577, Japan}
\affiliation{Faculty of Science, Hokkaido University, Kita 10 Nishi 8, Kita-ku, Sapporo 060-0810, Japan}
\email[show]{wada@astrophysics.jp}

\author[0000-0002-5488-1961]{Bert Vander Meulen}
\affiliation{European Space Agency (ESA), European Space Research and Technology Centre (ESTEC), Keplerlaan 1, 2201 AZ Noordwijk, The Netherlands}
\affiliation{Department of Physics and Astronomy, Universiteit Gent, Proeftuinstraat 86 N3, B-9000 Ghent, Belgium}
\email{bert.vandermeulen@ugent.be}

\begin{abstract}
We investigate the broad Br$\gamma$ line in the type~1 active galactic nucleus NGC~3783 using three-dimensional Monte Carlo electron-scattering radiative transfer with SKIRT. Paper~I reproduced the observed smooth core and broad wings by convolving an intrinsic Keplerian profile with the analytic electron-scattering kernel of \citet{BH2Laor2006}. Here we instead transport photons through explicit source and scattering geometries. For a partially covered Keplerian source, three-dimensional transfer broadens only the scattered photons and leaves an unscattered double-peaked core. \rev{For the thin Keplerian source motivated by Paper~I, a comparison of four scattering-geometry classes shows that neither a disk-like scatterer nor the tested halo variants reproduce both the core and wings.} We therefore construct a two-component-source (2CS) model containing an inner Keplerian source disk, a rotating disk haze, a nearly spherical halo haze, and direct emission from a slower outer disk. A reference 2CS calculation gives $\chidof=1.154$ at $i=15^\circ$. Polar cones with \rev{an input radial optical depth $\tau_{\rm hole}=0.1$} preserve the line profile and give a measured Thomson-thin sightline. The preferred low redistribution width, full-column ionization, and weak recombination emission cannot all be realized by a uniform thermal halo. The 2CS model is therefore a physically motivated parametric description whose medium must ultimately be tested with self-consistent radiative transfer, ionization, and dynamics.
\end{abstract}


\section{Introduction}\label{sec:intro}

Active galactic nuclei (AGNs) are powered by accretion onto supermassive black holes (SMBHs). Their characteristic broad emission lines originate in dense, rapidly moving gas close to the SMBH, conventionally called the broad-line region (BLR) \citep{BH2Peterson2006}. Determining the structure and kinematics of the BLR is essential both for precise SMBH mass measurements and for understanding how accretion is connected to AGN outflows. Because the BLR is so compact, however, the physical state, geometry, and kinematics of its emitting gas remain uncertain.

In the context relevant here, classical BLR pictures range from 
a very large population of discrete ionized clouds \citep{BH2Krolik1981,BH2Rees1989,BH2Baldwin1995} to continuous hydrodynamic structures such as rotating disks and disk winds \citep{BH2Murray1997,BH2Proga2000}. 
Dust-driven failed winds provide another proposed origin of the
low-ionization BLR: radiation pressure lifts dusty gas from the
accretion disk, while subsequent dust sublimation allows the gas
to fall back toward the disk
\citep{BH2Czerny2017}; see \citet{BH2Czerny2019} for a review
of BLR formation scenarios.
The absence of statistically detectable line-profile fluctuations from discrete clouds in high-quality spectra requires $N_{\rm cloud}\gtrsim10^6$--$10^8$ in cloud models \citep{BH2Arav1998,BH2Laor2006}, whereas a simple optically thin axisymmetric disk generally produces a double-peaked line, a morphology observed in only a minority of AGNs \citep{BH2Eracleous1994,BH2Strateva2003}.

Near-infrared interferometry with VLTI/GRAVITY has begun to resolve BLR scales directly \citep{BH2Gravity2018}. In NGC~3783, one of the brightest nearby type~1 AGNs, GRAVITY and SINFONI observations indicate that Br$\gamma$ arises on subparsec scales in a rotating, geometrically thick configuration \citep{BH2Gravity2021}. Nevertheless, a physical explanation for the detailed line profile remains incomplete.

\citet[hereafter Paper~I]{BH2Wada2026} coupled a three-dimensional radiation-hydrodynamic (RHD) simulation with CLOUDY \citep{BH2Ferland2017}. Br$\gamma$ emission arose near the illuminated surface of a thin rotating disk, at $T_e\sim10^4$~K and $n_e\sim10^8$--$10^{11}\,\mathrm{cm^{-3}}$. The intrinsic profile was too narrow and retained double-peaked substructure. Paper~I showed that electron scattering in diffuse ionized gas could broaden and smooth the line, using the approximate kernel of \citet{BH2Laor2006}. In that picture the intrinsic BLR is observed through an ionized ``haze.''

The broadening of Seyfert emission lines by electron scattering was first investigated by \citet{kaneko1968}, and subsequent work developed the interpretation of electron-scattering wings in active nuclei \citep{BH2Weymann1970,BH2Shields1981,BH2Laor2006}. The convolution used in Paper~I, however, reprocesses the entire line uniformly. It cannot represent the spatial distribution and velocity field of the scatterer, distinguish scattered from direct photons, or determine what fraction of the observed profile was actually redistributed.

We therefore replace the convolution with explicit three-dimensional Monte Carlo transport using SKIRT \citep{BH2Camps2015,BH2Camps2020,BH2VanderMeulen2023}. We ask three questions: (1) how should the low-optical-depth domain of the Laor analytic kernel be distinguished from the high-optical-depth numerical regime; (2) how does partial covering change the relation between scattering and convolution; and (3) which three-dimensional source--scatterer configuration reproduces the observed Br$\gamma$ profile? Section~\ref{sec:methods} describes the models and fitting procedure. Section~\ref{sec:results} presents the analytic-domain comparison, numerical convergence test, geometry \rev{comparison}, reference model, and component-ablation analysis. Section~\ref{sec:discussion} discusses the type~1 sightline, physical interpretation, and limitations. Common numerical settings and model nomenclature are collected in the Appendices.

\section{Methods and Models}\label{sec:methods}

\subsection{Radiative-transfer framework}

SKIRT propagates photon packets through arbitrary three-dimensional source and medium geometries, including bulk velocities and frequency redistribution \citep{BH2Camps2015,BH2Camps2020,BH2VanderMeulen2023}. We use SKIRT~9 in extinction-only mode and retain only Thomson scattering by free electrons. Electron velocities follow a Maxwellian distribution, so the transfer includes thermal Doppler redistribution but excludes absorption, re-emission, self-emission, polarization, and relativistic effects. For highly ionized gas at the inferred scattering-medium densities, $n_e\sim10^8\,\mathrm{cm^{-3}}$, and reference temperatures of $5\times10^3$--$10^4$~K, the free--free opacity at $2.166\,\mu$m is of order $10^{-4}$ of the Thomson opacity \citep{BH2Rybicki1979}, justifying its neglect in these scattering media (the separate ionization and self-emission constraints are discussed in Section~\ref{sec:selfemission}). At the inner source edge, $v_K/c=0.116$; the neglected transverse relativistic frequency correction scales as $(v_K/c)^2\simeq1.3\%$, while the characteristic beaming/asymmetry scale along the $i=15^\circ$ sightline (with inclination defined in Section~\ref{sec:fitproc}) is $(v_K/c)\sin i\simeq3\%$. Relativistic transfer may therefore modify the far-wing shape and red--blue asymmetry at the percent level.

The electron temperature sets the microscopic redistribution width,
\begin{equation}
\sigma_e=\left(\frac{kT_e}{m_e}\right)^{1/2}
=389\left(\frac{T_e}{10^4\ {\rm K}}\right)^{1/2}\kms,
\label{eq:sigmae}
\end{equation}
but not the Thomson scattering probability, which is set by optical depth and covering factor. Production spectra use $5\times10^6$ photon packets unless explicitly stated otherwise. \rev{The calculations testing the polar-cone geometry on a refined spatial grid (Section~\ref{sec:type1}) use $2\times10^6$ photon packets; the separate point-source diagnostic of the line-of-sight optical depth uses $10^5$ packets.}

For a specified detector direction and wavelength interval, we define the detected scattered-flux fraction as
\begin{equation}
\fscat=\frac{F_{\rm scat}}{F_{\rm scat}+F_{\rm dir}},
\end{equation}
where $F_{\rm scat}$ and $F_{\rm dir}$ are the band-integrated SKIRT-detected fluxes from photons that experienced at least one scattering and no scattering, respectively. Thus $\fscat$ is the scattered-flux fraction for one particular viewing direction, not the fraction of all emitted packets that scatter. It depends on geometry, covering factor, viewing angle, and wavelength interval. We reserve $F_{\rm trans}$ for the transparent (source) flux that would be detected if the medium were transparent; it includes the intrinsic line width and source kinematics but no scattering or attenuation.

\subsection{Br\texorpdfstring{$\gamma$}{gamma} source models}\label{sec:source}

The source present in every model, hereafter the \emph{inner source disk}, follows the geometry inferred in Paper~I. We measure the polar angle $\theta$ from the rotation axis. The source is a uniform wedge spanning $85^\circ\leq\theta\leq95^\circ$ at radii $10^{-4}$--$10^{-2}$~pc. Half of its luminosity is placed in 2000 emissivity clumps of radius $2\times10^{-4}$~pc and half in a smooth component. These clumps represent the nonuniform distribution of RHD cells selected by the photoionization calculation; they are not assumed to be self-gravitating clouds. \rev{The inner-source clumping prescription is inherited from Paper~I, whereas the outer source is smooth.}

The velocity field is Keplerian,
\begin{equation}
v_\phi(R)=3.49\times10^4
\left(\frac{R}{10^{-4}\ {\rm pc}}\right)^{-1/2}\kms,
\end{equation}
for $M_{\rm BH}=2.83\times10^7\msun$ \citep{BH2Gravity2021}. The source-line profile has a reference Gaussian dispersion $\sigma_v=100\kms$. This value defines a common baseline for the geometry and ablation experiments; it is not identified with the minimum of the separate $\sigma_v$ sensitivity sweep. The luminosity integrated over 2.10--2.24~$\mu$m is normalized to $10^{40}\,\mathrm{erg\,s^{-1}}$. Because the transfer equation is linear in the injected luminosity while the medium is held fixed, changing the common luminosity normalization changes only the fitted amplitude.

The selected two-component-source (2CS)  family adds a smooth \emph{outer source disk} with the same polar extent, $85^\circ\leq\theta\leq95^\circ$, from $10^{-2}$~pc to an outer radius of $3$ or $5\times10^{-2}$~pc. We define $f_{\rm out}=L_{{\rm Br}\gamma,{\rm out}}/L_{{\rm Br}\gamma,{\rm in}}$ as the ratio of the intrinsic 2.10--2.24~$\mu$m luminosities before transfer. The outer disk follows the same Keplerian law about the same SMBH, with $v_R=v_z=0$, and uses the same source-line dispersion as the inner disk. Its lower orbital speeds supply direct emission near line center; $f_{\rm out}$ is therefore not an observed-flux ratio.

\subsection{Scattering media and optical-depth normalization}\label{sec:tau}

Following Paper~I, we call the rotating wedge-like scatterer the \emph{disk haze} and a nearly spherical scatterer the \emph{halo haze}. We report one-sided electron-scattering depths: $\tau_e$ along the equatorial midplane for a disk or wedge, and $\tau_{\rm rad}$ from the central cavity to the outer boundary for a sphere or halo.\footnote{SKIRT's \texttt{OpticalDepthMaterialNormalization} is specified along the full model $X$ axis. For a smooth reflection-symmetric geometry, the reported one-sided depth is half the SKIRT input normalization: $\tau_e$ for a disk or wedge and $\tau_{\rm rad}$ for a sphere or halo. For a clumpy medium, $\tau_e$ denotes \rev{half the input optical-depth normalization}; the actual source-to-escape depth varies among sightlines and emission sites.} Extended sources in every geometry therefore sample a distribution of escape depths rather than a single central value.

The independent haze wedge and the disk and halo hazes in the reference 2CS model are smooth; the co-spatial, source-shaped scatterer tested in Section~\ref{sec:doublepeak} is explicitly clumpy.

The disk haze corotates with the Keplerian source and has $T_d=10^4$~K. Halo models are uniform, centrally concentrated as $r^{-2}$, or broken power laws, and are either stationary in bulk or rotating. In the reference model, SKIRT samples a single Maxwellian electron distribution at $T_h=5\times10^3$~K. Earlier sensitivity runs encoded \rev{an additional microscopic velocity dispersion} through an equivalent ElectronMix temperature, $T_{\rm eff}=T_h+(m_e/k)\sigma_{\rm mic}^2$, \rev{where $\sigma_{\rm mic}$ is the equivalent one-dimensional dispersion of an additional isotropic Gaussian electron-velocity component. This contribution was represented by the effective temperature rather than by an independent turbulent velocity field.} They therefore constrain the total microscopic electron-speed distribution but cannot distinguish thermal motion from an independent isotropic Gaussian microscopic contribution. Spatially correlated or macroscopic turbulent support would require a different three-dimensional medium velocity model.

The spatial mesh is an adaptive octree with maximum levels 8--13. Reconstructed full-axis depths for the smooth closed geometries converge to their analytic inputs to better than 1\%. \rev{For the polar-cone model, the optical depth measured along the central sightline differs from the input radial optical depth inside the cones because of the discretized cone boundary (Section~\ref{sec:type1}).}

\begin{deluxetable*}{>{\raggedright\arraybackslash}p{0.15\textwidth}>{\raggedright\arraybackslash}p{0.38\textwidth}>{\raggedright\arraybackslash}p{0.37\textwidth}}
\tablecaption{Adopted Components and Explored Parameters\label{tab:params}}
\tabletypesize{\footnotesize}
\tablehead{\colhead{Component} & \colhead{Assumed Geometry} & \colhead{Main Parameters}}
\startdata
Inner source disk & Clumpy Keplerian wedge, $85^\circ\leq\theta\leq95^\circ$, $10^{-4}$--$10^{-2}$~pc & $\sigma_v=50$--$500\kms$; $L=10^{40}\,\mathrm{erg\,s^{-1}}$ \\
Outer source disk & Smooth Keplerian wedge, $85^\circ\leq\theta\leq95^\circ$, $10^{-2}$~pc--$r_{\rm out}$ & $f_{\rm out}=0.03$--0.30; $r_{\rm out}=0.03$ or $0.05$~pc \\
Disk haze & Corotating Keplerian wedge, $10^{-4}$--$10^{-2}$~pc, $T_d=10^4$~K & polar extent from $85^\circ\leq\theta\leq95^\circ$ to $60^\circ\leq\theta\leq120^\circ$; reference is the latter, $\tau_e=3$ \\
Halo haze & Spherical shell, $10^{-4}$--$10^{-2}$~pc: uniform, $r^{-2}$, or broken power law & $\tau_{\rm rad}=0.3$--8; reference $T_h=5\times10^3$~K; reported one-parameter minimum at $2\times10^3$~K \\
\enddata
\end{deluxetable*}

\pagebreak
\subsection{Data and fitting procedure}\label{sec:fitproc}

We compare the models with the continuum-normalized SINFONI Br$\gamma$ profile of NGC~3783 \citep{BH2Gravity2021}, after subtraction of the narrow-line-region component. We adopt the same distance of 38.5~Mpc as \citet{BH2Gravity2021}. Inclination is measured from the rotation axis, so $i=0^\circ$ is face-on. The standard detector grid is $i=0^\circ,10^\circ,15^\circ,30^\circ,45^\circ,$ and $90^\circ$; selected 2CS sensitivity calculations add $5^\circ,20^\circ,$ and $25^\circ$. All quoted $\chidof$, dip/peak ratios, and \rev{full widths at half maximum (FWHM)} are measured after convolution with a Gaussian line-spread function (LSF) of resolving power $R\simeq4000$, or $\sigma_{\rm inst}\simeq32\kms$. 
We distinguish the instrumental dispersion $\sigma_{\rm inst}$
from the local intrinsic source-line dispersion $\sigma_v$;
FWHM denotes the width of the final transferred and
LSF-convolved profile.
We define dip/peak as the intensity at $v=0$ divided by the maximum intensity within $|v|<3500\kms$; unity denotes a centrally peaked profile, while a value below unity measures the residual central depression.

The observational error supplied with the reduced spectrum is a constant $5.23\times10^{-3}$ in continuum-normalized flux per spectral point. We fit $N=727$ wavelength points over 2.100--2.278~$\mu$m with three nuisance parameters: amplitude, continuum offset, and velocity shift. Thus $k=3$ and $\nu=724$. All three-dimensional transfer models and the Paper~I-style Laor convolution use the same fitting procedure. The count $k=3$ does not include the effective degrees of freedom introduced by selecting geometry, optical depth, temperature, inclination, and source structure from discrete grids.

\section{Results}\label{sec:results}

\subsection{Laor-kernel domain and high-optical-depth convergence}\label{sec:kernel}

The analytic electron-scattering kernel of \citet{BH2Laor2006} gives the exponential e-folding width
\begin{equation}
\sigma=1.1\,\sigma_e\left[\ln\left(\tau_e^{-1}\right)\right]^{-0.45}.
\label{eq:laor}
\end{equation}
\rev{Here $\sigma_e$ is the one-dimensional electron thermal velocity dispersion defined in Equation~(\ref{eq:sigmae}).}
The Laor calculation assumes isotropic illumination and electron velocities but does not solve spatial transport. The profile at each successive scattering order is obtained by convolving the preceding profile, and its flux is weighted recursively by $\tau_e$; the contributions from all scattering orders are then added. We refer to this summed profile as the scattering-order sum. Equation~(\ref{eq:laor}) is an analytic fit to its exponential-wing scale. 
The high-optical-depth convergence tests described below use
a clumpy, disk-like scattering medium with the same spatial
distribution as the source.
\rev{The equatorial optical-depth normalization $\tau_e$} therefore does not
represent the range of optical depths encountered by photons
emitted at different locations and traveling in different
directions. This sequence cannot directly validate the
Laor wing-width relation at a matched optical depth.

\begin{figure}
\centering
\includegraphics[width=0.92\linewidth]{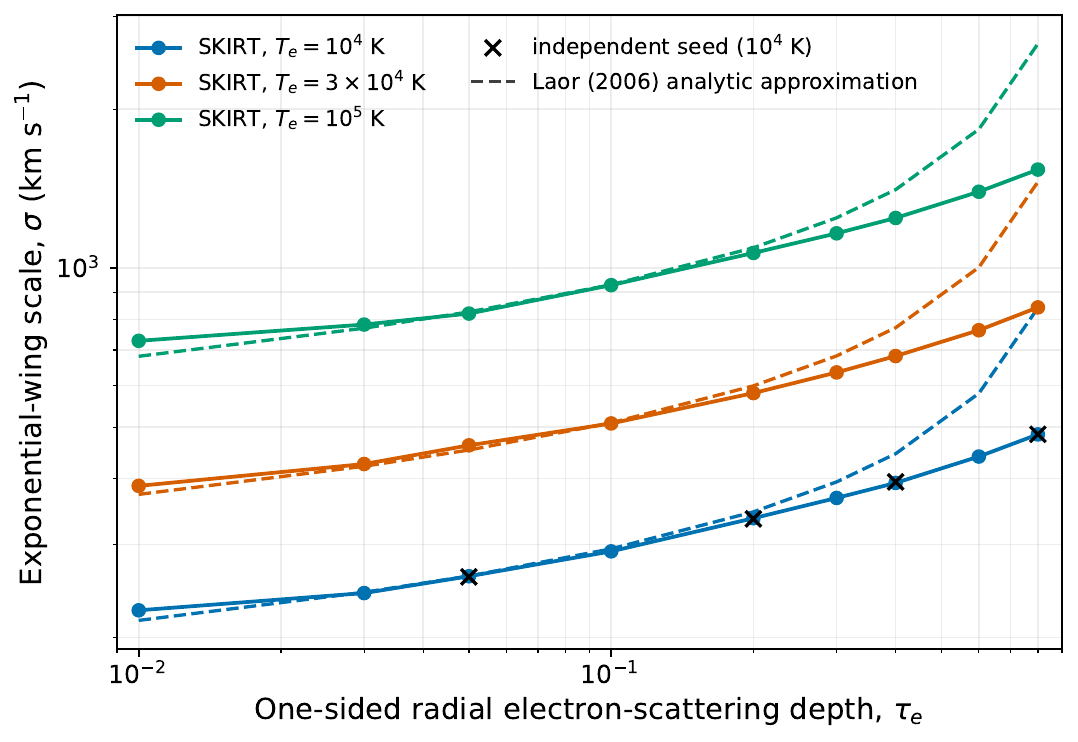}
\caption{Definition-controlled electron-scattering comparison. Solid curves and circles show exponential-wing scales from a static uniform SKIRT sphere whose one-sided radial depth $\tau_{\rm rad}$ is matched to the analytic $\tau_e$; dashed curves show the Laor analytic approximation in equation~(\ref{eq:laor}), fitted to the profile obtained by summing all scattering-order contributions. Crosses repeat four $10^4$~K calculations with an independent random seed. Agreement at low depth gives way to a systematic difference as geometry-dependent escape alters the scattering-order distribution.}
\label{fig:kernel}
\end{figure}

We therefore constructed a spatial benchmark with an isotropic central point source, a smooth static spherical shell, and a one-dimensional spherical grid. The one-sided radial depth $\tau_{\rm rad}$ was set equal to the analytic $\tau_e$; independently, $-\ln(F_{\rm dir}/F_{\rm trans})$ recovers every reported depth to the stated precision. We used a narrow $50\kms$ source line and fitted both scattered wings between $10^{-5}$ and $10^{-2}$ of their peak. For $\tau_e=0.03$--0.2 and $T_e=10^4$--$10^5$~K, the SKIRT e-folding widths agree with equation~(\ref{eq:laor}) to within 3.1\% (Figure~\ref{fig:kernel}). At $\tau_e=0.01$ the very weak wings differ by up to 7\%. At $\tau_e=0.3$, 0.4, 0.6, and 0.8, the temperature-independent SKIRT/Laor ratios decline to approximately 0.93, 0.88, 0.76, and 0.58. Independent random seeds at four representative depths change the measured widths by less than 0.5\%, so the trend is not Monte Carlo noise.

The growing difference is expected because a physical sphere gives each scattered photon a direction- and position-dependent remaining escape depth, whereas the Laor order sum assigns a fixed factor $\tau_e$ to every additional scattering. Thus the comparison validates the low-depth redistribution scale while also delimiting where the zero-dimensional order-weighting approximation ceases to represent spatial transfer. We do not extrapolate the Laor kernel to the optically thick models below.

We separately tested numerical behavior in the multiple-scattering regime using a narrow source line, zero source bulk velocity, and a source-shaped clumpy wedge \rev{with equatorial optical-depth normalizations $\tau_e=1,2,4,$ and 7}. In every run, the scattering medium is a \texttt{ClumpyGeometryDecorator} realization of the same $5^\circ$ torus envelope as the source, with 50\% of the medium mass assigned to 2000 clumps. Across this sequence, the scattered-component dispersion increases monotonically from 684 to $1624\kms$, while the flux fraction at $|v|=1500$--$6000\kms$ rises from 0.10 to 0.38. \rev{At an equatorial optical-depth normalization of $\tau_e=7$}, increasing the packet count from $2\times10^7$ to $5\times10^7$ changes the FWHM, wing fraction, and total dispersion by less than 0.2\%. This establishes packet-number convergence and monotonic broadening for this high-depth numerical sequence, not agreement with an independent analytic solution.

\subsection{Why three-dimensional scattering is not a convolution}\label{sec:doublepeak}

\begin{figure*}
\centering
\includegraphics[width=0.96\textwidth]{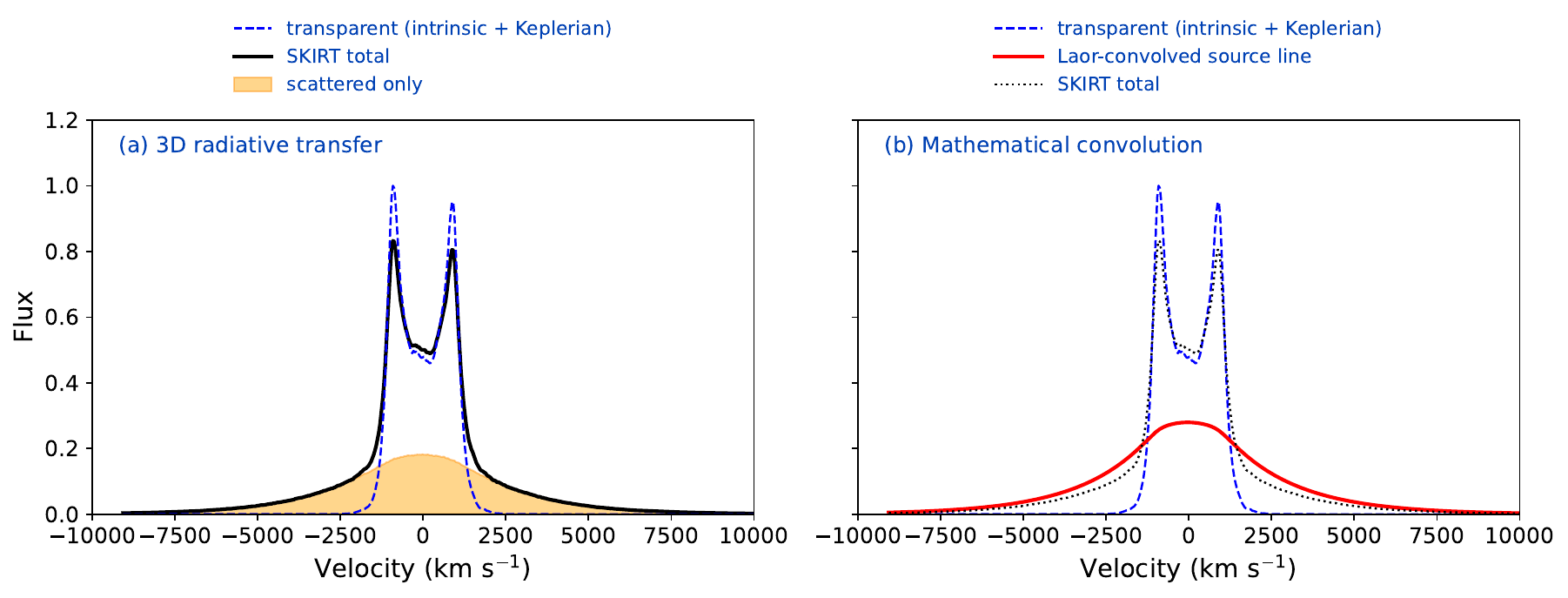}
\caption{\rev{Three-dimensional scattering compared with convolution for a partially covering, source-shaped clumpy wedge with an equatorial optical-depth normalization of $\tau_e=8$, $T_e=10^5$~K, and $i=15^\circ$. The blue dashed ``transparent (intrinsic + Keplerian)'' curve is identical in both panels. Every curve is divided by the peak flux of that transparent spectrum, preserving relative amplitudes. (a) Three-dimensional radiative transfer: ``SKIRT total'' (black solid) is broadened but remains double-peaked (dip/peak $=0.601$); the orange filled region, ``scattered only,'' is the detected scattered flux (band-integrated $\fscat=0.490$). (b) Mathematical convolution: ``Laor-convolved source line'' (red) uses the symmetric exponential (Laplace) kernel of scale $2080\kms$ adopted in Paper~I and has a single peak. ``SKIRT total'' (black dotted) repeats exactly the total spectrum in panel (a) for comparison.}}
\label{fig:convolution}
\end{figure*}

When Keplerian rotation is restored, three-dimensional scattering with a redistribution width comparable to the symmetric exponential (Laplace) kernel with scale $2080\kms$ adopted in Paper~I broadens the wings but does not erase the double-peaked core (Figure~\ref{fig:convolution}). At $i=15^\circ$, the SKIRT profile has dip/peak $=0.601$, whereas convolving the same source-line profile produces a single peak (dip/peak $=1.000$).

The co-spatial, source-shaped scatterer is a clumpy wedge with the same polar extent as the source, $85^\circ\leq\theta\leq95^\circ$, \rev{and an equatorial optical-depth normalization of $\tau_e=8$}. This higher-depth realization gives a band-integrated detected-flux fraction $\fscat=0.490$ over 2.10--2.24~$\mu$m at $i=15^\circ$, making the scattering-induced broadening directly visible while preserving the source and covering geometry. The total profile remains double-peaked (dip/peak $=0.601$), whereas a convolution redistributes the entire source profile and becomes single-peaked.

\begin{figure}
\centering
\includegraphics[width=0.86\linewidth]{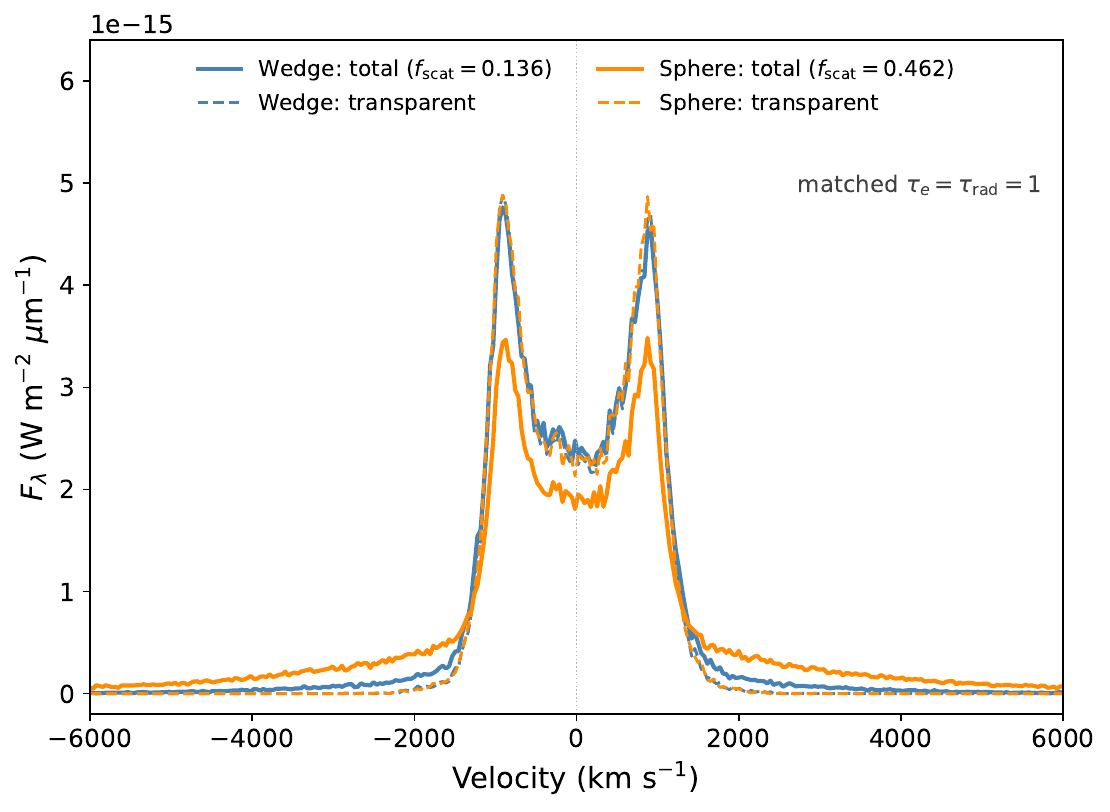}
\caption{Effect of scatterer geometry at matched one-sided optical depth, $\tau_e=\tau_{\rm rad}=1$, viewed at $i=15^\circ$. Solid and dashed curves show total and transparent (source) spectra for smooth wedge (blue) and spherical (orange) scatterers. Both have $T_e=10^5$~K, the same Keplerian velocity field and radial limits, and the same clumpy source; only the scatterer envelope differs. Their band-integrated detected scattered-flux fractions are 0.136 and 0.462. These diagnostic calculations use $10^5$ photon packets each.}
\label{fig:covering}
\end{figure}

The covering dependence is isolated in Figure~\ref{fig:covering} at $i=15^\circ$. The source, smooth-medium density law, radial limits, electron temperature, Keplerian velocity field, one-sided depth ($\tau_e=1$ for the wedge and $\tau_{\rm rad}=1$ for the sphere), and detector are identical; only the medium envelope changes from a thin wedge ($85^\circ\leq\theta\leq95^\circ$) to a sphere. In the 2.10--2.24~$\mu$m band, the detected scattered-flux fraction rises from $\fscat=0.136$ to 0.462. Thus even at matched one-sided depth, covering geometry strongly changes the detected scattered fraction. The absolute value also remains directional: scattering both removes photons from and redirects photons into the detector.

\subsection{Systematic exploration of scattering geometries}\label{sec:geometry}

\begin{figure*}
\centering
\includegraphics[width=0.88\textwidth]{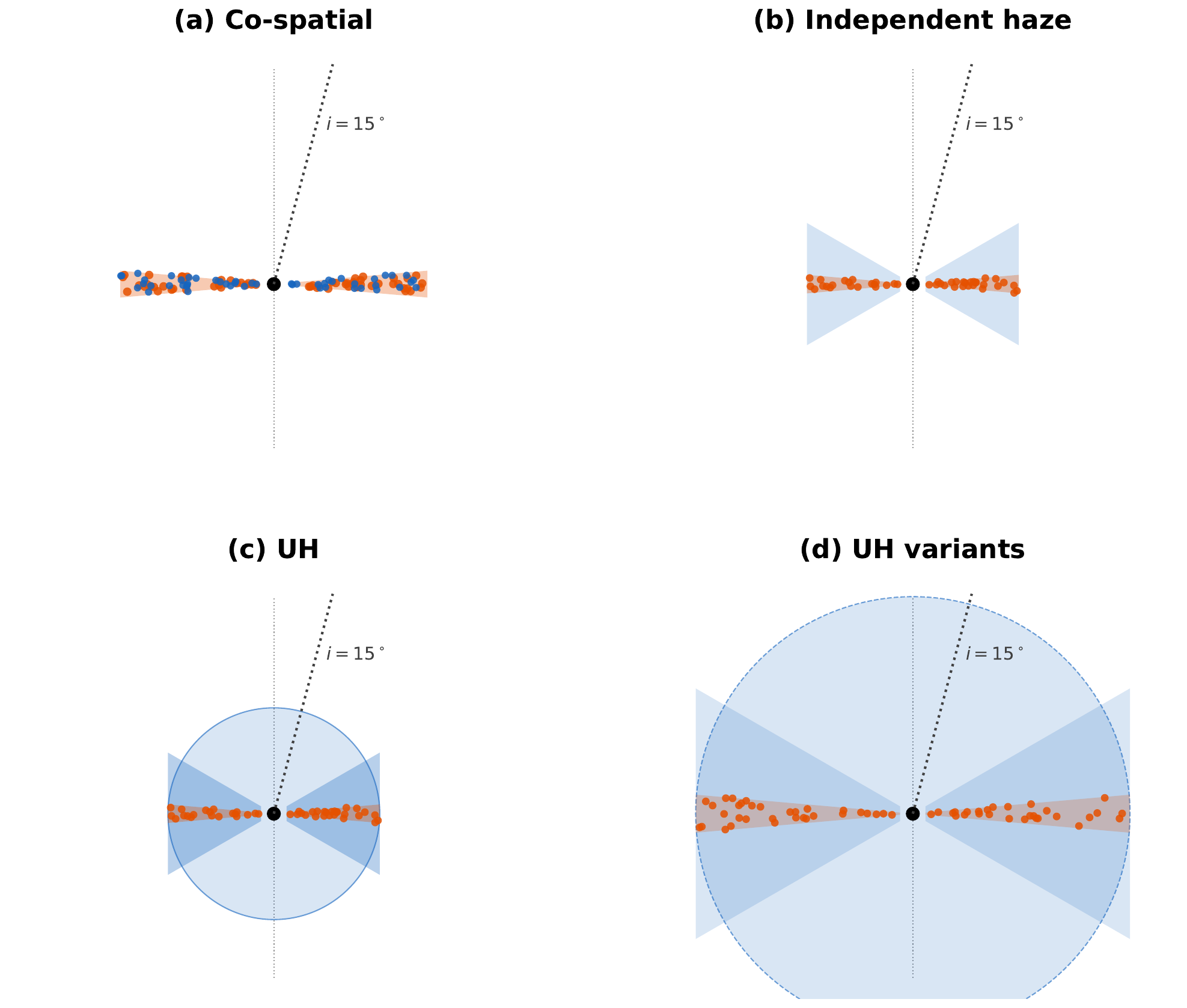}
\caption{\rev{Schematic $r$--$z$ sections of four broad geometry classes, not to scale. Orange denotes Br$\gamma$ emitters and blue denotes electron scatterers: (a) a co-spatial source-shaped clumpy scatterer; (b) an independent smooth haze wedge; (c) a uniform halo (UH) haze together with a disk haze; and (d) variants of UH, with an extended rotating halo (ERH) shown as the representative example. The centrally concentrated halo (CCH) and broken-power-law halo (BPLH) differ from ERH in their radial density and velocity prescriptions and are described in Section~\ref{sec:geometry}.}}
\label{fig:families}
\end{figure*}

The observed Br$\gamma$ line is macroscopically smooth and nearly single-peaked, with broad wings and FWHM of order $4000\kms$, although a shallow central depression remains under the adopted LSF and error model. \rev{We compare four scattering-geometry classes in Figure~\ref{fig:families}.} Within each class, models were normalized primarily at fixed one-sided equatorial depth $\tau_e$ (or $\tau_{\rm rad}$ for a smooth spherical model). These residual comparisons therefore apply to models matched in reported one-sided optical depth, not to models matched in total electron mass or in the emission-weighted optical depth from photon production sites to escape.

\paragraph{(a) Co-spatial scatterer shaped like the source.}
The high-depth, partially covering clumpy disk used above produces visible scattering wings but retains a double-peaked total profile, with band-integrated detected $\fscat=0.490$ at $i=15^\circ$ \rev{for an equatorial optical-depth normalization of $\tau_e=8$}.

\paragraph{(b) Independent wedge (Haze).}
Separating the scatterer from the clumpy source and widening its polar extent from $75^\circ\leq\theta\leq105^\circ$ to $45^\circ\leq\theta\leq135^\circ$ raises the covering factor and changes $\sigma_{\rm scat}$ from about 650 to $1050\kms$. The double-peaked direct component nevertheless survives.

\paragraph{(c) Disk haze plus uniform halo haze (UH).}
A uniform halo distributes electron column throughout the emitting volume and, together with the rotating disk haze, gives the lowest residual among the single-source models. UH gives $\chidof=1.316$ at $i=15^\circ$, while its minimum on the standard inclination grid is 1.257 at $i=10^\circ$. Its dip/peak ratio saturates near 0.9, so scattering alone still does not completely fill the central depression.

\paragraph{(d) Variants of UH.}
Figure~\ref{fig:families}(d) shows the extended rotating halo (ERH) as a representative of the tested UH variants. Extending a uniformly rotating halo to 0.1~pc makes the core single-peaked but produces a profile that is too narrow; the lowest-residual ERH case has $\chidof=3.111$. A centrally concentrated $r^{-2}$ halo (CCH) performs worse at fixed reported one-sided depth because most of the electron column lies inside the region traversed by few photons, reducing the detected $\fscat$; its lowest residual is $\chidof=3.077$. A broken-power-law halo (BPLH), with $\rho\propto r^{-2}$ inside a break and $r^{-4}$ outside, behaves similarly; varying its break radius has little effect because the column remains concentrated near the center, and its lowest residual is $\chidof=4.845$. \rev{Thus, concentrating electrons near the center does not ensure efficient scattering of the extended source: substantial electron columns must lie along the photons' escape paths.} The detailed distinctions among ERH, CCH, and BPLH are stated here rather than drawn as separate panels.

The common limitation is that changing only the scatterer either preserves the double peak or obtains a single peak at the cost of an incorrect width. We therefore adopt UH as the single-source baseline and add a distinct low-velocity emitter.

\begin{deluxetable*}{>{\raggedright\arraybackslash}p{0.11\textwidth}c>{\raggedright\arraybackslash}p{0.30\textwidth}>{\raggedright\arraybackslash}p{0.39\textwidth}}
\tablecaption{Geometry \rev{Comparison} and Principal Outcomes\label{tab:models}}
\tabletypesize{\footnotesize}
\tablehead{\colhead{Model} & \colhead{Panel} & \colhead{Scattering Geometry} & \colhead{Outcome}}
\startdata
Co-spatial & (a) & Source-shaped clumpy disk & Partial covering; double peak survives \\
Haze & (b) & Independent smooth wedge & $\sigma_{\rm scat}\sim650$--$1050\kms$; double peak survives \\
UH & (c) & Uniform halo haze plus disk haze & $\chidof=1.316$ at $i=15^\circ$; lowest-residual single-source model \\
UH variants & (d) & ERH, CCH, and BPLH (see Section~\ref{sec:geometry}) & Lowest $\chidof=3.077$--$4.845$ across the three variants \\
2CS & \rev{Figs.~\ref{fig:bestfit}, \ref{fig:concept2cs}} & UH plus outer source disk & Reference $\chidof=1.154$; separate one-parameter minima 1.118 and 1.128
\enddata
\tablecomments{\rev{Panels (a)--(d) refer to Figure~\ref{fig:families}; the 2CS spectrum and geometry are shown in Figures~\ref{fig:bestfit} and \ref{fig:concept2cs}, respectively.}}
\end{deluxetable*}

\subsection{Two-component-source reference model}\label{sec:bestfit}

\begin{figure*}
\centering
\includegraphics[width=0.88\textwidth]{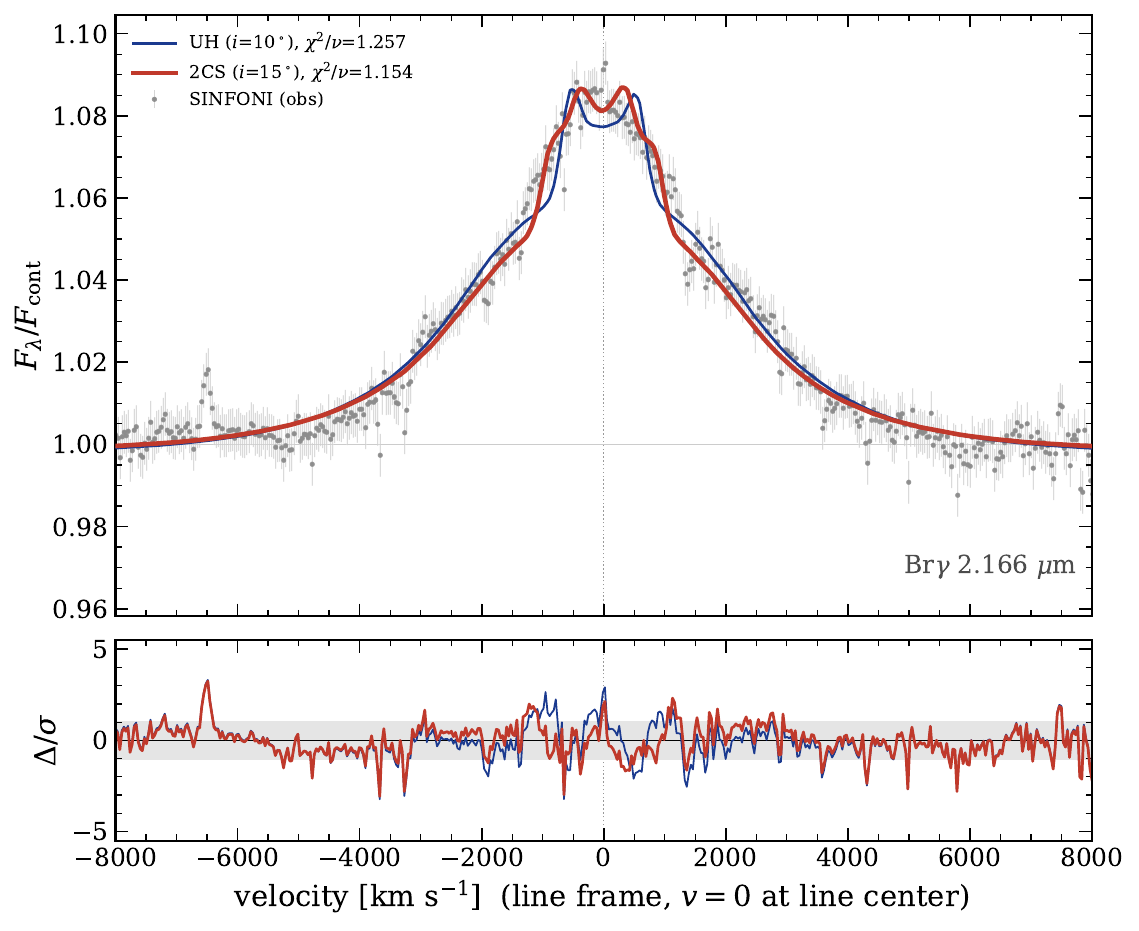}
\caption{\rev{SINFONI Br$\gamma$ profile and residuals. The thick red solid curve is the reference two-component-source (2CS) model illustrated in Figure~\ref{fig:concept2cs}, with $f_{\rm out}=0.05$ and $i=15^\circ$ ($\chidof=1.154$). The thinner blue solid curve is the uniform-halo (UH) configuration at its minimum-grid inclination, $i=10^\circ$ ($\chidof=1.257$). The narrow feature near $-6500\kms$ is consistent with H$_2$ 1--0 S(1), also identified in the SINFONI spectrum by \citet{BH2Gravity2021}, and is not included in the Br$\gamma$ models. Masking $-6900<v<-6100\kms$ in the displayed velocity frame and refitting the three nuisance parameters gives $\chidof=1.135$ (2CS) and 1.236 (UH), each with $\nu=700$, without changing their ordering or fitted velocity shifts; the normalizations change by less than 0.5\%. The plotted fits and quoted main-text statistics retain the original unmasked wavelength range.}}
\label{fig:bestfit}
\end{figure*}

\begin{figure*}
\centering
\includegraphics[width=0.90\textwidth]{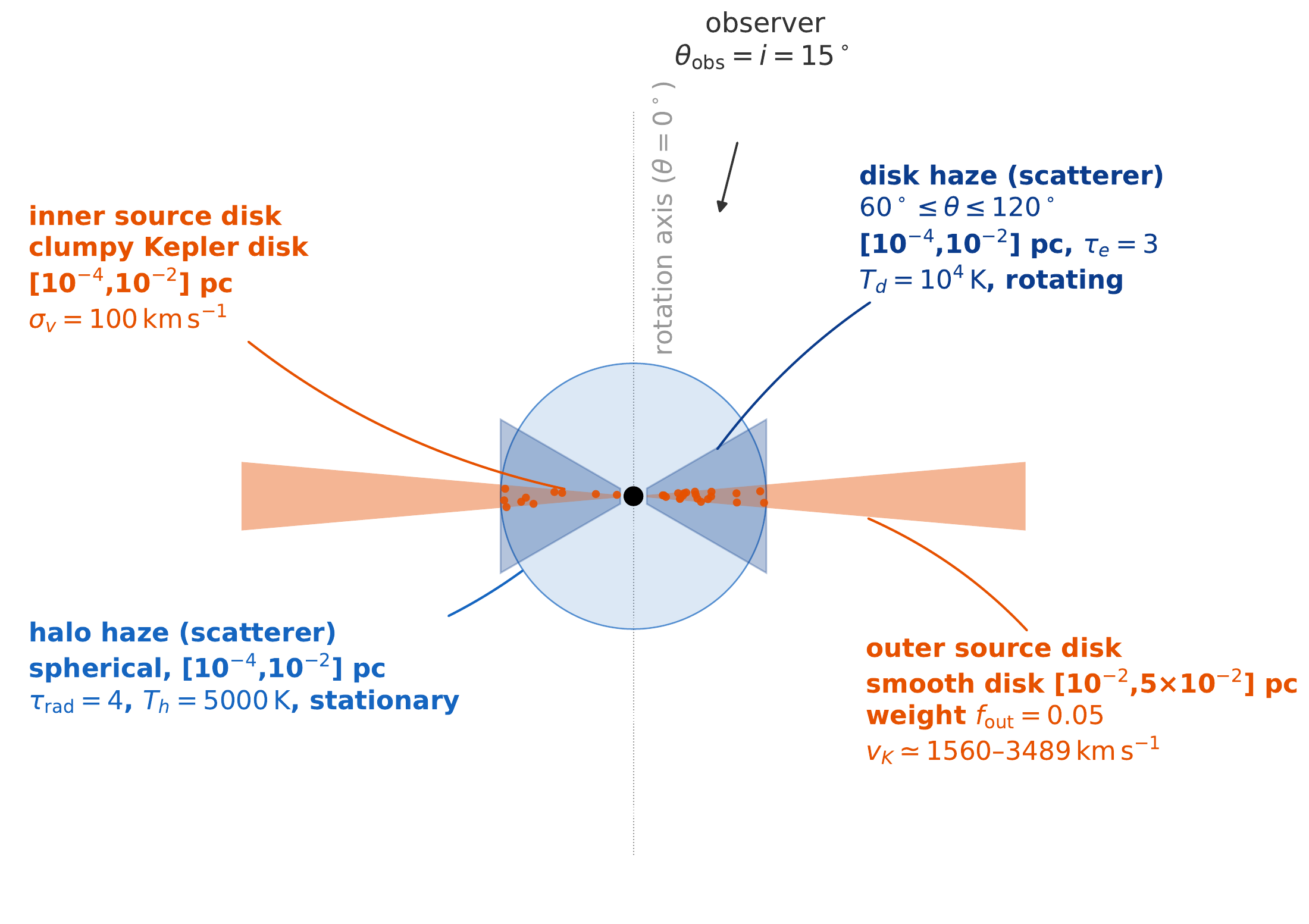}
\caption{Schematic $r$--$z$ section of the reference 2CS model, not to scale. All polar angles $\theta$ are measured from the rotation axis: the observer is at $\theta_{\rm obs}=i=15^\circ$, and the rotating disk haze spans $60^\circ\leq\theta\leq120^\circ$. Orange marks the inner and outer Br$\gamma$ source disks; dark-blue hatching is the disk haze with one-sided equatorial depth $\tau_e=3$; and light blue is the nearly spherical halo haze with $\tau_{\rm rad}=4$ and $T_h=5\times10^3$~K. Direct outer-disk emission fills the central depression. The polar-cone variant 2CS-p20 is introduced separately in Section~\ref{sec:type1}.}
\label{fig:concept2cs}
\end{figure*}

\rev{Figure~\ref{fig:bestfit} compares the two-component-source (2CS) reference profile with the observations, and Figure~\ref{fig:concept2cs} illustrates its source and scattering geometry. Because scattering alone stalls at dip/peak $\sim0.9$ within the tested thin-source models, 2CS adds direct emission from slower outer gas.} The scattering medium is the UH configuration: a uniform nearly spherical halo with $\tau_{\rm rad}=4$ and $T_h=5\times10^3$~K plus a disk haze with midplane radial depth $\tau_e=3$. The inner source spans $10^{-4}$--$10^{-2}$~pc. \rev{Electron scattering of its photons by the disk and halo hazes forms the broad wings. The smooth outer source lies outside the scatterer and fills the shallow central depression with essentially direct emission; its intrinsic luminosity is 5\% of the inner-source luminosity.} Its orbital speeds are roughly 1500--$3500\kms$, corresponding to projected speeds of about 400--$900\kms$ at $i=15^\circ$.

\rev{We compare two representative outer-source radii to test the velocity width of this direct core contribution. The compact variant, 2CS-c, shares the 2CS geometry but ends at $3\times10^{-2}$~pc and gives $\chidof=1.196$. Extending the outer edge to $5\times10^{-2}$~pc lowers the minimum orbital speed from about 2000 to $1600\kms$ (projected speeds of about 520 and $400\kms$ at $i=15^\circ$), supplying more low-velocity emission to fill the central depression and giving $\chidof=1.154$. These radii sample compact and more extended core-emitting configurations; they are not a confidence interval on the outer radius, and the residual difference does not exclude 2CS-c.} We designate this latter configuration, with $T_h=5\times10^3$~K, intrinsic $\sigma_v=100\kms$, $f_{\rm out}=0.05$, and $i=15^\circ$, as the \emph{reference model}. It is the common baseline for figures, component ablations, and one-parameter comparisons, rather than a claim of a joint physical best fit. Relative to that baseline, UH at the same inclination gives $1.316$, and the Paper~I-style Laor convolution evaluated through the same fitting pipeline gives $1.202$.

Holding all other reference parameters fixed, the halo-temperature sweep has its minimum at $T_h=2\times10^3$~K with $\chidof=1.118$, whereas the independent source-line-dispersion sweep has its minimum at $\sigma_v=200\kms$ with $\chidof=1.128$. These are separate one-parameter sensitivity minima and have not been combined into a jointly optimized model. Both also expose physical tensions if interpreted literally: $2\times10^3$~K is below the expected equilibrium temperature of a strongly photoionized plasma, while a local $200\kms$ source-line dispersion is far above the hydrogen thermal width and has no dynamical origin in the present parametric source model.

\rev{We also tested the sensitivity to inner-source clumping. Replacing the clumpy inner source with a smooth disk, while retaining the other reference-model parameters, gives $\chidof=1.155$ and dip/peak $=0.935$, compared with 1.154 and 0.934 for the reference model. Thus, the line-profile fit is insensitive to the adopted inner-source clumping prescription in this test.}

\subsection{Component ablation}\label{sec:ablation}
\begin{figure*}
\centering
\includegraphics[width=0.98\textwidth]{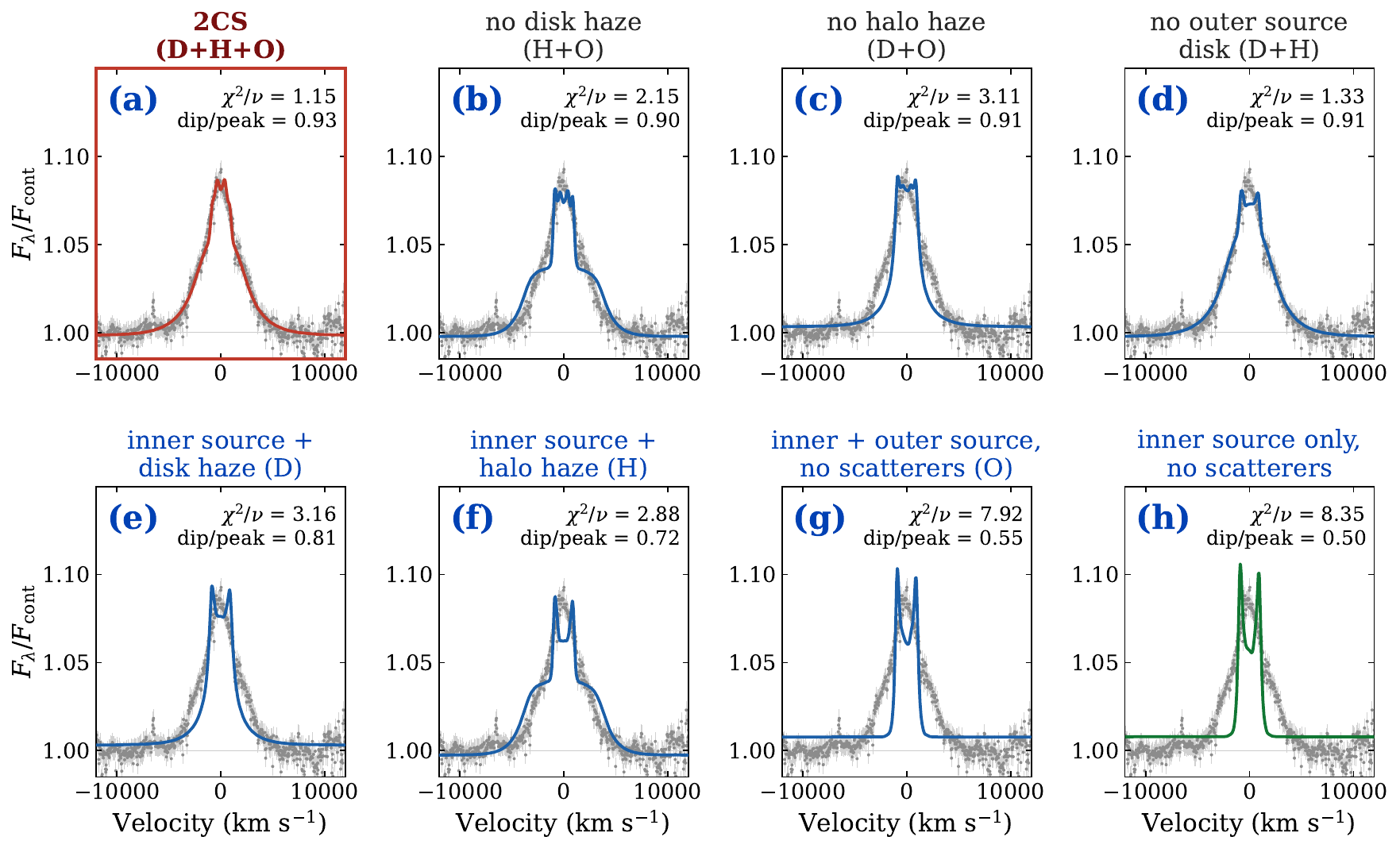}
\caption{Component-ablation experiments starting from the reference 2CS model in panel (a). Each panel compares the reference model or one of its ablations with SINFONI at $i=15^\circ$. The disk haze provides central broadening and partial horn filling, the halo produces most of the broad exponential wings, and the outer source fills the dip between the horns. \rev{The inner source disk is present in every panel; the letters D, H, and O in parentheses denote the retained disk haze, halo haze, and outer source disk, respectively. Panel (g) retains both source disks without scatterers; panel (h) retains only the inner source without scatterers. Both profiles include the instrumental line-spread function. Table~\ref{tab:ablation} lists all eight cases and their panel labels.}}
\label{fig:ablation}
\end{figure*}

Starting from the reference 2CS model illustrated in
Figure~\ref{fig:concept2cs}, we remove the disk haze (D),
halo haze (H), and outer source disk (O), \rev{individually and in combinations, including all three,} while retaining the inner source disk.
\rev{Figure~\ref{fig:ablation} and Table~\ref{tab:ablation} separate three roles. The rotating disk haze broadens the inner-source profile locally and smooths its Keplerian horns, as shown by its removal and isolation in panels (b) and (e). The halo supplies most of the broad scattered wings (panels (c) and (f)); removing it is more damaging than removing the disk haze, and removing both scatterers produces a strong double peak in panel (g), with $\chidof=7.918$. The outer source supplies low-projected-velocity direct flux that fills the residual central dip (panels (d) and (g)): removing it changes dip/peak from 0.934 to 0.906 and changes $\chidof$ from 1.154 to 1.331. Panel (h), which removes D, H, and O, isolates the inner source without scatterers and gives $\chidof=8.350$. These experiments identify the roles of the components within the adopted 2CS model, with the other parameters held fixed. Alternative source geometries and velocity fields are discussed in Section~\ref{sec:predictions}.}

\begin{deluxetable*}{c>{\raggedright\arraybackslash}p{0.24\textwidth}cc>{\raggedright\arraybackslash}p{0.34\textwidth}}
\tablecaption{Component Ablation at $i=15^\circ$\label{tab:ablation}}
\tabletypesize{\footnotesize}
\tablehead{\colhead{\rev{Fig.~\ref{fig:ablation}}} & \colhead{Removed} & \colhead{Retained} & \colhead{$\chidof$} & \colhead{Interpretation}}
\startdata
\rev{(a)} & None & D+H+O & 1.154 & Reference configuration \\
\rev{(b)} & Disk haze & H+O & 2.150 & Inner broadening and horn smoothing weaken \\
\rev{(c)} & Halo haze & D+O & 3.111 & Broad scattered wings are lost \\
\rev{(d)} & Outer source & D+H & 1.331 & Central depression returns \\
\rev{(e)} & \rev{Halo haze + outer source} & D & 3.156 & Central broadening; partial horn filling \\
\rev{(f)} & \rev{Disk haze + outer source} & H & 2.880 & Most broadening; broad exponential wings \\
\rev{(g)} & \rev{Disk haze + halo haze} & O & 7.918 & Direct flux fills the inter-horn dip \\
\rev{(h)} & \rev{Disk haze + halo haze + outer source} & \rev{None} & \rev{8.350} & \rev{Inner source only; unscattered double peak}
\enddata
\tablecomments{D, H, and O denote disk haze, halo haze, and outer source disk. \rev{The panel labels correspond to Figure~\ref{fig:ablation}. The inner source disk is retained in every case; the Retained column lists only D, H, and O. Cases (g) and (h) use the transparent spectra of the two-source and inner-source configurations, respectively.} All values include the instrumental LSF and use $\nu=724$.}
\end{deluxetable*}

\section{Discussion}\label{sec:discussion}

\subsection{Revisiting the Paper I picture}

Paper~I reproduced the smooth core and broad wings by applying the Laor electron-scattering kernel to the entire intrinsic profile. The three-dimensional calculation shows that this treatment overestimates core smoothing when the source is only partially covered. Real scattering redistributes only the detected fraction $\fscat$, leaving direct Keplerian structure in the rest of the line. Across the model families explored, including clumpy realizations drawn from axisymmetric envelope geometries, broad wings still arise naturally from electron scattering, but a separate low-velocity direct component is needed to fill the center. Model 2CS therefore refines, rather than rejects, the ``BLR behind haze'' interpretation.

\subsection{Type~1 consistency: polar cones to \texorpdfstring{$\theta=20^\circ$}{theta = 20 degrees}}\label{sec:type1}

If the reference halo were literally uniform and spherical with $\tau_{\rm rad}=4$, only $e^{-4}=0.018$ of the central continuum would escape directly, apparently contradicting the type~1 classification. We therefore computed model 2CS-p20, in which the input density is reduced inside two polar cones, $\theta\leq20^\circ$ and $\theta\geq160^\circ$, \rev{with an input one-sided radial optical depth inside the cones of $\tau_{\rm hole}=0.1$}, while the halo between the cones, disk haze, and both sources remain unchanged. A boundary at $\theta=15^\circ$ was avoided because the adopted $i=15^\circ$ sightline would coincide with the numerical density discontinuity.

At $i=15^\circ$, matched $2\times10^6$-packet high-resolution octree runs of the spherical and 2CS-p20 geometries give $\chi^2=835.7$ and 836.7, respectively, for $\nu=724$ ($\chidof=1.154$ and 1.156). The matched-grid difference is $\Delta\chi^2=+1.0$ for 2CS-p20, so the two models are indistinguishable for spectral fitting at this precision; 2CS-p20 is a type~1 consistency test rather than a fit improvement. Its LSF-convolved profile has FWHM $=3502\kms$, dip/peak $=0.931$, and a band-integrated detected $\fscat=0.871$. The two cones occupy $1-\cos20^\circ=0.060$ of the full solid angle and reduce the halo mass by about 5.9\%, so the high-covering multiple-scattering region remains nearly intact.

A central point-source diagnostic through the complete 2CS-p20
model at $i=15^\circ$ gives \rev{a line-of-sight optical depth $\tau_{\rm los}\equiv-\ln(F_{\rm dir}/F_{\rm trans})\simeq0.158$},
a direct fraction $F_{\rm dir}/F_{\rm trans}\simeq0.853$,
and $N_e\simeq2.4\times10^{23}\,\mathrm{cm^{-2}}$,
well below the closed-halo radial column of
$\sim6\times10^{24}\,\mathrm{cm^{-2}}$.
Thus the full three-dimensional transport yields a
Thomson-thin central sightline.\footnote{
\rev{The measured $\tau_{\rm los}$}
differs from \rev{the input radial optical depth $\tau_{\rm hole}=0.1$} because
octree cells reconstruct the sharp cone boundary; the
$i=15^\circ$ sightline lies only $5^\circ$ inside it.
Both depths are well below unity. This central-source
direct fraction is distinct from the extended-line
scattered fraction $\fscat$ quoted above.}
The polar cones preserve the Br$\gamma$ profile but remain
a phenomenological prescription rather than a
self-consistent ionization structure.

\subsection{Possible physical origins of the components}\label{sec:origins}

The four components of 2CS are parametric structures introduced to explain the line profile. Their line-formation roles are quantified, but their physical identities are not uniquely determined. At $r\sim10^{-2}$~pc, the dynamical time is about three years, so a thick configuration requires support or continued replenishment.

\paragraph{Inner source disk and disk haze.}
The inner source can be identified with the photoionized recombination layer selected in the RHD+CLOUDY calculation of Paper~I. Its outer radius agrees with the H$\beta$ reverberation radius of about 10 light days \citep{BH2Onken2002} and the GRAVITY radius of about 0.013~pc \citep{BH2Gravity2021}. The disk haze may be the more highly ionized atmosphere above the same emitting layer. Its characteristic density, estimated from the radial midplane depth $\tau_e=3$, is $n_e\sim1.5\times10^8\,\mathrm{cm^{-3}}$. Its smaller wedge volume reduces its emission measure (EM) relative to the spherical halo discussed below, but its recombination emission has not been calculated self-consistently; we assume here that it is sufficiently ionized to act primarily as a scatterer.

\paragraph{Nearly spherical, bulk-stationary halo haze.}
For $\tau_{\rm rad}=4$ and $r_{\rm out}=10^{-2}$~pc, a uniform estimate gives $n_e\sim2\times10^8\,\mathrm{cm^{-3}}$ and a total mass of about $20\msun$. \rev{Adding microscopic velocity dispersions of $\sigma_{\rm mic}=500$, 1000, and $2000\kms$ in quadrature through the effective temperature} changes $\chidof$ to 1.409, 1.969, and 2.928. As explained in Section~\ref{sec:tau}, these runs alter a single Maxwellian $T_{\rm eff}$; they do not add an independent turbulent field. Thermal motion and an independent isotropic Gaussian microscopic contribution with the same total variance would be spectroscopically degenerate in this implementation, so lowering $T_h$ while holding that total variance fixed would generate the same redistribution rather than an independent physical test. Coherent or spatially correlated macroturbulence, which could contribute to structural support, requires a separate medium velocity model.

Possible alternatives include a slow thermal or photoevaporative wind, a continuously replenished transient fountain, and magnetic support. A $\sim10\kms$ slow wind would require a spherical mass flux of order $0.06\msun\,\mathrm{yr^{-1}}$, comparable to the estimated accretion rate $\sim0.05\msun\,\mathrm{yr^{-1}}$. A radiation-driven fountain could yield a quasi-spherical time-averaged distribution \citep{BH2Wada2012,BH2Wada2026}, but the microscopic electron-speed distribution must obey the spectral constraint. Magnetic pressure can provide geometrical support without directly requiring a large isotropic thermal or turbulent speed. Hyper-magnetized disk calculations likewise predict a thick, multiphase structure extending from torus scales toward the SMBH \citep{BH2Hopkins2025,BH2Hopkins2026}. None of these possibilities alone resolves the mismatch between the $T_h=2\times10^3$~K one-parameter line-shape minimum (or the $5\times10^3$~K reference width) and a photoionized equilibrium temperature of order $2\times10^4$~K.

\paragraph{Outer source disk.}
The 0.01--0.05~pc source lies between the main BLR and the dust sublimation radius. If its 5\% luminosity is produced by a uniform Case~B recombination emitter, the required density is of order $10^7\,\mathrm{cm^{-3}}$, intermediate between typical BLR and narrow-line-region densities. It could represent an intermediate-line region \citep{BH2Brotherton1994}, the ionized inner edge of the torus, returning fountain gas, or the cooler outer phase of a continuous magnetized disk.

These components may therefore be different phases of one circulation structure rather than separate objects: a recombination layer, its ionized atmosphere, a diffuse or magnetically elevated scattering phase, and an outer settling layer. Testing this interpretation requires direct transfer through self-consistent RHD or magnetohydrodynamic (MHD) density, temperature, velocity, and ionization fields. A residual narrow-line component after the adopted subtraction could also contribute flux between the Keplerian horns; quantifying this possible alternative to part of the outer-source contribution is deferred to future work.

\subsection{Halo self-emission and ionization consistency}\label{sec:selfemission}

Interpreting the halo as a uniform, fully ionized, volume-filling sphere creates an internal inconsistency. Its emission measure is fixed by $\tau_{\rm rad}=4$ and $r_{\rm out}=10^{-2}$~pc at ${\rm EM}\simeq4.7\times10^{66}\,\mathrm{cm^{-3}}$. Case~B then gives
\begin{equation}
L_{\rm halo}^{\rm Br\gamma}\simeq1.6\times10^{40}
\left(\frac{T}{10^4\ {\rm K}}\right)^{-0.9}\mathrm{erg\,s^{-1}}
\end{equation}
\citep{BH2Hummer1987}, comparable to the adopted inner-source luminosity at photoionized temperatures. A stationary halo-emission source with intrinsic $\sigma_v=50\kms$ remains a narrow spike with FWHM $\simeq230\kms$ after transfer and LSF convolution. It cannot replace the smooth low-velocity outer source while retaining the narrow reference redistribution.

Hotter electrons reduce recombination emission but broaden the frequency redistribution. Moreover, the total column $N_e\simeq6\times10^{24}\,\mathrm{cm^{-2}}$ need not remain fully ionized through its depth merely because the illuminated face has a high ionization parameter. We have not found a uniform solution that simultaneously gives narrow redistribution, full-column ionization, and weak self-emission. The halo should therefore be regarded as an effective representation of an inhomogeneous medium until scattering and recombination emission are solved self-consistently.

\subsection{Comparison with GRAVITY and observable tests}\label{sec:predictions}

The reference $i\simeq15^\circ$, source/scatterer scale near $10^{-2}$~pc, and adopted $M_{\rm BH}=2.83\times10^7\msun$ are broadly consistent with the GRAVITY rotating-thick-disk fit: mean radius $R_{\rm BLR}\simeq0.013$~pc, inclination $i\simeq23^\circ$, and position angle $\simeq295^\circ$ \citep{BH2Gravity2021}. The two models encode different physics, so a direct comparison of their $\chidof$ values is not meaningful. A future SKIRT calculation of differential phases and velocity-dependent photocenters would permit a quantitative interferometric test.

\rev{Our thin source disk is physically motivated by Paper~I: the RHD calculation coupled to photoionization modeling identified gas satisfying the BLR emission conditions near the illuminated surface of a thin, approximately Keplerian disk. The present work examines how electron scattering modifies the emission from this structure. The two approaches are complementary: GRAVITY constrains the source geometry and kinematics directly from the observed line profile and differential phases \citep{BH2Gravity2021}, while our model starts from a source structure derived from RHD and photoionization calculations and explores the effects of scattering on its emergent spectrum.}

\rev{Magnetic support, which was not included in Paper~I, provides a plausible alternative route to a geometrically thick emitting disk. Magnetically dominated accretion models predict thick, multiphase structures containing BLR-emitting gas \citep{BH2Hopkins2025}. The persistence of a strongly magnetically supported thick disk remains debated. The non-cosmological ideal-MHD simulations of \citet{BH2Guo2026}, without explicit magnetic diffusivity, yield a dense thin midplane plus a diffuse magnetized layer when the thermal scale height is resolved, qualitatively resembling our emitting disk plus haze in density geometry. Whether such a structure develops the substantial population of radial-motion-dominated orbits suggested by GRAVITY remains an open question. Nevertheless, a thick emitting configuration with substantial radial motions could produce a smooth, centrally filled line profile resembling that produced by a thinner source viewed through an electron-scattering haze. This is a possible degeneracy of the spatially unresolved spectrum, rather than an equivalence demonstrated by our calculations.}

\rev{Within the adopted 2CS model, the halo mainly supplies the broad wings, while the disk haze smooths the inner Keplerian horns and direct outer-source emission fills the remaining central depression (Figure~\ref{fig:ablation}). A different intrinsic source geometry and velocity field could change the relative contributions required from these components. Establishing the connection between magnetic support, the emitting-gas kinematics, and the emergent line profile requires future MHD calculations coupled to photoionization and radiative transfer, tested against interferometric and velocity-resolved reverberation constraints.}

Because the Thomson cross section is effectively wavelength-independent across these optical and near-infrared lines, the same scatterer should produce related heavy wings in other transitions, while their source emissivities may differ. The single-epoch analysis of NGC~3783 by \citet{BH2Kuhn2024} found smooth, heavy-tailed profiles in H$\alpha$, H$\beta$, H$\gamma$, Pa$\beta$, and He~\textsc{i}, with much of the line-to-line variation explained by different radial emissivity distributions in a common geometry. Keeping the 2CS scatterer fixed while replacing the source emissivity for each transition is therefore a strong test.

Finally, scattered photons travel farther than direct photons. A scale of $10^{-2}$~pc corresponds to a geometric light-travel time of about 12 days, but the $\tau_{\rm rad}=4$ halo can produce a broader delay distribution through multiple scattering. The outer direct source also has its own geometric response. The model therefore predicts a velocity-resolved transfer function, but its wing--core lag must be calculated rather than identified with a single light-crossing time. A related technique has located an equatorial scattering region in Mrk~6 from the delay of polarized broad-line emission \citep{BH2Shablovinskaya2020}. Future SKIRT runs using the same three-dimensional geometry can enable polarized transport, as implemented and tested for scattering calculations by \citet{BH2Peest2017} and \citet{BH2VanderMeulen2024}, and record photon path lengths to predict wavelength-dependent Stokes spectra and light-travel-time distributions. The current SKIRT implementation can record photon arrival-time lags determined by geometric path lengths.\footnote{\url{https://skirt.ugent.be/root/_user_time_lags.html}} These capabilities were not enabled in the present calculation and require no assumption that either observable follows a single characteristic scale.

\section{Summary and Conclusions}\label{sec:conclusion}

We used three-dimensional Monte Carlo electron-scattering transfer to examine and extend the convolution approximation used for NGC~3783 in Paper~I. Our principal conclusions are:
\begin{enumerate}
\item A definition-controlled static spherical SKIRT calculation agrees with the Laor exponential-wing scale to within 3.1\% at $\tau_e=0.03$--0.2, but becomes systematically narrower at larger depth because spatial escape changes the scattering-order distribution. \rev{Additional tests with a clumpy scattering medium having the same geometry as the source show that the line-width measures converge as the photon-packet count increases and that the scattered component broadens as the optical depth increases.} In a partially covered geometry, only scattered photons are redistributed, so broad wings can form while an unscattered Keplerian double peak survives.
\item \rev{A comparison of four scattering-geometry classes} shows that a thin disk or wedge retains the central double peak, while the tested UH (uniform halo) variants either produce profiles that are too narrow or concentrate the electron column where few photons traverse it. At fixed input depth, the most effective scattering geometry places high covering and electron column where the photon flux is large.
\item \rev{Within the thin-source model family motivated by Paper~I, the following component decomposition provides a successful reference profile but is not unique among alternative source geometries and kinematics.} Adding an outer-disk intrinsic luminosity equal to 5\% of the inner-source luminosity to a disk haze plus halo haze gives the reference 2CS (two-component-source) calculation, $\chidof=1.154$ at $i=15^\circ$. Separate one-parameter sweeps reach $\chidof=1.118$ at $T_h=2\times10^3$~K and 1.128 at intrinsic $\sigma_v=200\kms$. Ablations identify disk-haze inner broadening and horn smoothing, halo-generated broad wings, and outer-source filling of the central dip.
\item Microscopic thermal and isotropic Gaussian nonthermal speeds
are degenerate when represented by a single effective
electron temperature; macroscopic turbulent support
requires a different velocity model.
\item A uniform halo cannot simultaneously satisfy the preferred redistribution width, full-column ionization, and weak self-emission; this inconsistency requires a multiphase or otherwise nonuniform scattering medium.
\end{enumerate}

The immediate next steps are to solve electron scattering, recombination emission, and continuum transfer through self-consistent RHD/MHD and photoionization structures; and test a single scatterer against multiple emission lines, reverberation delays, interferometric observables, and polarization.

\begin{acknowledgments}
We thank the anonymous referee for constructive comments that helped clarify the physical interpretation and improve the presentation of this work.
We thank the SKIRT development team and the teams responsible for the GRAVITY and SINFONI observations of NGC~3783. K.W. acknowledges support from JSPS KAKENHI Grant Number 25H00671. B.V.M. acknowledges support through the European Space Agency (ESA) Research Fellowship Programme in Space Science and thanks Ghent University and SRON for their hospitality throughout this work. 
OpenAI Codex and Anthropic Claude were used to assist with manuscript editing, revision, and figure preparation; 
the authors take responsibility for the scientific content and final text.
\end{acknowledgments}

\facilities{VLT:Yepun (SINFONI)}

\software{SKIRT \citep{BH2Camps2015,BH2Camps2020,BH2VanderMeulen2023}, CLOUDY \citep{BH2Ferland2017}}

\clearpage
\appendix

\section{Numerical Settings and Implementation}\label{sec:settings}

\begin{deluxetable}{ll}[h]
\tablecaption{Baseline SKIRT Settings\label{tab:settings}}
\tablehead{\colhead{Item} & \colhead{Setting}}
\startdata
Code & SKIRT~9 \\
Mode & \texttt{ExtinctionOnly}; no absorption or re-emission \\
Medium & \texttt{ElectronMix}; Thomson scattering; Maxwellian velocities; polarization off \\
Normalization & \texttt{OpticalDepthMaterialNormalization} along equatorial $X$ axis \\
Source geometry & \texttt{TorusGeometry} plus \texttt{ClumpyGeometryDecorator} \\
Velocity field & \texttt{CylindricalVectorField}; $v_\phi\propto R^{-1/2}$ \\
Intrinsic line & \texttt{FileSED}; reference $\sigma_v=100\kms$ \\
Detector & \texttt{SEDInstrument}; transparent/direct/scattered separation on \\
Wavelength grid & 1.80--2.55 $\mu$m; 1200 points \\
Spatial grid & Octree, maximum level 8--13 \\
Photon packets & Normally $5\times10^6$; see Section~\ref{sec:type1} for diagnostic exceptions \\
Distance & 38.5 Mpc \\
SMBH mass & $2.83\times10^7\msun$
\enddata
\end{deluxetable}

\section{Model Nomenclature and Reference Configuration}\label{sec:names}

\begin{deluxetable}{ll}
\tablecaption{Model Tags\label{tab:names}}
\tablehead{\colhead{Tag} & \colhead{Meaning}}
\startdata
Co-spatial & Source-shaped clumpy scattering disk \\
Haze & Smooth wedge independent of the source \\
CCH & Centrally concentrated $\rho\propto r^{-2}$ halo \\
UH & Uniform halo plus disk haze; single-source baseline \\
ERH & Uniform rotating halo extended to $0.1$~pc \\
BPLH & Broken-power-law halo: $r^{-2}$ inner and $r^{-4}$ outer \\
2CS-c & Compact outer source ending at $0.03$~pc \\
2CS & Reference outer source ending at $0.05$~pc \\
2CS-p20 & 2CS with polar cones $\theta\leq20^\circ$ and $\theta\geq160^\circ$, \rev{and input radial optical depth $\tau_{\rm hole}=0.1$}
\enddata
\end{deluxetable}

The reference 2CS model at $i=15^\circ$ contains rotating inner and outer source wedges spanning $85^\circ\leq\theta\leq95^\circ$, at $10^{-4}$--$10^{-2}$~pc and $10^{-2}$--$5\times10^{-2}$~pc, respectively, with intrinsic luminosity ratio $f_{\rm out}=0.05$; a rotating disk haze at $10^{-4}$--$10^{-2}$~pc spanning $60^\circ\leq\theta\leq120^\circ$, with $\tau_e=3$ and $T_d=10^4$~K; and a stationary uniform spherical halo haze at $10^{-4}$--$10^{-2}$~pc with $\tau_{\rm rad}=4$ and $T_h=5\times10^3$~K. The inner edge is a common central cavity, the outer source begins at the outer boundary of both scattering media, and the media therefore do not cover the outer source. In 2CS-p20, the halo density is reduced inside $\theta\leq20^\circ$ and $\theta\geq160^\circ$ \rev{using an input radial optical depth $\tau_{\rm hole}=0.1$}; the halo between the cones and the disk haze are unchanged.

\bibliography{references}
\bibliographystyle{aasjournalv7.1}

\end{document}